\documentclass[12pt]{article}
\pdfoutput=1
\usepackage{amsmath,amsfonts,amscd,amssymb,cite}
\usepackage[cp1251]{inputenc}
\usepackage[english]{babel}
\usepackage[T2A]{fontenc}
\usepackage{upref}
\usepackage{xspace}
\usepackage{enumerate}
\usepackage{amsthm}
\usepackage{amsxtra}
\usepackage{array}
\usepackage[pdftex]{graphicx}
\usepackage{xcolor}
\definecolor{linkcolor}{HTML}{FF0000} 
\definecolor{urlcolor}{HTML}{0000CC} 
\usepackage{hyperref}
\hypersetup{pdfstartview=FitH,  linkcolor=linkcolor, urlcolor=urlcolor, colorlinks=true}

\let\cal\mathcal

\begin{document}
\title {
$${}$$
{\bf
Non-free gas of dipoles of non-singular screw dislocations and the shear modulus near the melting
}}
\author{
$${}$$
{\bf\Large Cyril Malyshev}\\
$${}$$\\
{\it\small Steklov Institute of Mathematics}
{\it\small (St.-Petersburg Department)}\\
{\it\small Fontanka 27, St.-Petersburg, 191023, RUSSIA}}

\date{}

\maketitle

\def \bbe{\boldsymbol\be}
\def \bchi{\boldsymbol\chi}
\def \bdl{\boldsymbol\dl}
\def \bphi{\boldsymbol\phi}
\def \bsi{\boldsymbol\si}
\def \bta{\boldsymbol\eta}
\def \bxi{\boldsymbol\xi}
\def \bna{\boldsymbol\nabla}
\def \bvphi{\boldsymbol\varphi}
\def \bPsi{\boldsymbol\Psi}
\def \binc{\boldsymbol\Inc}
\def \bx{\bf{x}}
\def \bQ{{\bf Q}}
\def \bcP{\boldsymbol{\cal P}}

\def \al{\alpha}
\def \be{\beta}
\def \ga{\gamma}
\def \dl{\delta}
\def \ze{\zeta}
\def \nb{\nabla}
\def \th{\theta}
\def \la{\lambda}
\def \si{\sigma}
\def \om{\omega}
\def \z{\zeta}
\def \Ga{\Gamma}
\def \Dl{\Delta}
\def \La{\Lambda}
\def \Si{\Sigma}
\def \Ph{\Phi}
\def \Om{\Omega}
\def \ph{\varphi}
\def \vt{\vartheta}
\def \ep{\varepsilon}
\def \ka{\varkappa}

\def \cA{\cal A}
\def \cB{\cal B}
\def \cC{\cal C}
\def \cD{\cal D}
\def \cE{\cal E}
\def \cF{\cal F}
\def \cG{\cal G}
\def \cI{\cal I}
\def \cJ{\cal J}
\def \cK{\cal K}
\def \cN{\cal N}
\def \cP{\cal P}
\def \cR{\cal R}
\def \cS{\cal S}
\def \cT{\cal T}
\def \cY{\cal Y}
\def \cZ{\cal Z}
\def \cM{{\cal M}}
\def \cL{{\cal L}}
\def \CU{{\cal U}}
\def \CW{{\cal W}}

\def \BC{\mathbb{C}}
\def \BD{\mathbb{D}}
\def \BZ{\mathbb{Z}}
\def \BR{\mathbb{R}}
\def \BQ{\mathbb{Q}}
\def \at{{\rm arctan}\,}
\def \ch{{\rm ch}\,}
\def \sh{{\rm sh}\,}
\def \th{{\rm th}\,}
\def \bg{{\rm bg}\,}
\def \Tr{{\rm Tr}\,}
\def \tr{{\rm tr}\,}
\def \Det{{\rm Det}\,}
\def \Inc{{\rm Inc}\,}
\def \diag{{\rm diag}\,}
\def \e{{\rm e}\,}
\def \c{{\rm c}\,}
\def \m{{\rm m}\,}
\def \d{{\rm d}}
\def \o{{\rm o}\,}

\def \w{\widetilde}
\def \h{\widehat}
\def \nt{{\widetilde n}}
\def \t{\times}
\def \r{\rangle}
\def \l{\langle}
\def \lav{\l\!\l}
\def \rav{\r\!\r}
\def \ld{\ldots}
\def \IM{\Im}
\def \RE{\Re}
\def \1{^{-1}}
\def \cd{\partial}

\vskip1.0cm

\begin{abstract}
\noindent The behavior of the shear modulus caused by proliferation of dipoles of non-singular screw dislocations with finite-sized core is considered. The representation of two-dimensional Coulomb gas with smoothed-out coupling is used, and the stress--stress correlation function is calculated. A co\-n\-vo\-lut\-ion integral expressed in terms of the modified Bessel function $K_0$ is derived in order to obtain the shear modulus in approximation of interacting dipoles. Implications are demonstrated for the shear modulus near the melting transition which are due to the singularityless character of the dislocations.
\end{abstract}


\thispagestyle{empty}
\newpage

\section{Introduction}
\label{sec1}

The physics of nanotubes, nanowires and
graphene sheets is of importance as far as development of modern technologies is
concerned \cite{saito, toman, gul}.
Dislocations as imperfections of the
crystalline ordering have attracted
appreciable attention from the viewpoint of real properties of nanostructures
\cite{voz1, dkl, aif1, shuo, voz2, graph1, yaz, g5, g6, akat}. For instance, the multilayer
nanotubes can contain within their walls
screw dislocations lying along the tube
axis \cite{g5, g6}.
It is also of interest that the mechanical and electronic properties of the graphene sheets can significantly be influenced by dislocations \cite{voz1, voz2, graph1, yaz}.

Dislocations are of importance also for  two-dimensional melting
\cite{holz, nel1, nel12, nel13} being an example of the defect-mediated phase transitions in the two-dimensional systems
\cite{ber1, kos1, pop1, pop2}. The textbooks \cite{kl11, kl12} summarize a large body of original work on the gauge theory of the line-like defects and the phase transitions caused by their proliferation. In turn, the multivalued fields are of great importance in the condensed matter physics to describe defects and the corresponding phase transitions \cite{mult}. Multivaluedness of appropriate transformation functions (of the displacement field in the case of dislocated crystals) is responsible for the topological
non-triviality of the line-like defects. The multivalued coordinate transformations relate flat spaces to general affine spaces with curvature and torsion which are relevant to the theory of dislocations and disclinations \cite{kl12, mult}.

The stress tensor of conventional dislocation is singular on its line since the dislocation core is not captured by the standard elasticity theory \cite{hirth}. The dislocation solutions characterized by elimination of the axial singularities have been investigated by means of the non-local elasticity \cite{cem1}, the gradient elasticity \cite{ga, g1, lma1}, and the gauge approaches \cite{val, ed1, mal, laz2}. The modification of the conventional stresses occurs due to additional contributions arising so that the singularities are smoothed out within the finite-sized cores. The non-local elasticity and the gradient elasticity are the generalized continuum theories which effectively take into account interatomic forces for
explanation of the material behavior on the scales of the defect cores \cite{lma1}. Although the Lagrangian gauging \cite{val, ed1, mal, laz2} has a different grounding than the generalized elasticity \cite{cem1, ga, g1, lma1}, the corresponding solutions arising in the both approaches are in mutual agreement.
Since the sizes of the dislocation cores are comparable with the characteristic scales of nanotubes, effects due to the non-conventional stresses should be valuable for nanostructures.

Proliferation of the
dislocation dipoles is responsible for the renormalization of the elastic constants \cite{holz, nel1, nel12, nel13, kos1, rab}. According to \cite{wigner}, the type of the defect mediated two-dimensional melting transition is determined by the value of the higher-gradient elastic constant called the angular stiffness (see also \cite{mult}). Therefore it is attractive from the viewpoint of nano-physics to study the behavior of the elastic moduli
using the modified dislocation solutions
\cite{cem1, ga, g1, lma1, val, ed1, mal, laz2}.

Thermodynamical description of collection of the non-singular screw dislocations at non-zero temperature is equivalent to that of two-dimensional Coulomb gas with smoothed out coupling \cite{mal3}.
In the present paper we calculate the shear modulus $\mu_{\rm ren}$ influenced by the interacting dipoles in the Coulomb-like system describing the modified dislocations. Implications for $\mu_{\rm ren}$ of the short-ranged correlations characterizing the cores are demonstrated, including the properties of $\mu_{\rm ren}$ near the melting transition. It should be stressed that the given approach is concerned with individual defects but not with an appropriate field theory of the type of the disorder field theories developed in \cite{kl11, kl12, mult} for various phase transitions (e.g., superfluidity, superconductivity, melting).

The paper is organized as follows. Section~1 is the introductory section.
The partition
function describing the modified screw dislocations as a Coulomb-like system is given in section~2 in the dipole approximation. The stress--stress correlation function is obtained in Section~2 in the case of interacting dipoles. A convolution integral over infinite plane expressed through the modified Bessel function $K_0$ and its derivatives is obtained in Section~3 which allows us to calculate the multi-dipole contributions into the stress--stress correlation function.
In Section~4 the shear modulus is obtained, and its modification caused by the singularityless character of the dislocations is demonstrated. Discussion in Section~5 closes the paper.

\section{The partition function and the stress--stress correlation function}
\label{sec2}

Let us begin with the stress field ${\si}_{i}({\bf x})$ of straight screw dislocation with smoothed-out singularity lying within an elastic body, \cite{cem1, ga, g1, lma1, val, ed1, mal, laz2}. Non-singularity of the dislocation implies that ${\si}_{i}({\bf x})$ is given by superposition of two terms, ${\si}_{i}({\bf x})={\si}^{\rm b}_{i}({\bf
x})+{\si}^{\rm c}_{i}({\bf x})$, $i=1, 2$, where we abbreviate:
${\si}^{\#}_{i}\equiv {\si}^{\#}_{i 3}$
($\#$ is $\rm b$ or $\rm c$).
The dislocation line is parallel to the axis $Ox_3$ of Cartesian frame, and ${\bf x}\equiv (x_1, x_2)$.
The term ${\si}^{\rm b}_{i}$ is the conventional long-ranged stress of singular dislocation, while ${\si}^{\rm c}_{i}$ describes the modification of ${\si}^{\rm
b}_{i}$ within the core. In the framework of (anti-)plane elasticity, the \textit{modified screw dislocation} located at ${\bf y}$ and possessing the Burgers vector ${\bf b}\upuparrows Ox_3$ is characterized by superposition of the stresses ${\si}^{\#}_{i}({\bf x}) = \mu \epsilon_{i k}\,\cd_{{x}_k} f^{\#}({\bf x})$ ($\epsilon_{1 2}=-\epsilon_{2 1}=1$, $\#$ is ${\rm b}$ or ${\rm c}$), where the stress potentials are $f^{\rm b}({\bf x})=\frac{-b}{2\pi}\log|{\bf x}-{\bf y}|$ or $f^{\rm c}({\bf x})=\frac{-b}{2\pi} K_0(\kappa|{\bf x}-{\bf y}|)$. Here $b\equiv |{\bf b}|$, $\mu$ is the shear modulus,  $\kappa\equiv(\mu/\ell)^{\frac12}$ is the dimensionless parameter given by
$\mu$ and by the scale of the dislocation core energy $\ell$, and
$K_0(s)$ is the modified Bessel function.

Let us begin with a model system given by infinite elastic cylinder containing $2\cN$ non-singular screw dislocations with unit Burgers vectors ${\bf b}_{\widetilde I}$, $1\le {\widetilde I}\le 2\cN$, lying along its axis $O x_3$. Under the condition of ``electro-neutrality'' the number of positive dislocations (${\bf b}_{\widetilde I} \upuparrows Ox_3$) intersecting the plane $x_1 O x_2$ at the points $\{{\bf
y}^+_I\}_{1\le I\le\cN}$ is equal to the number of negative ones (${\bf b}_{\widetilde I} \downdownarrows Ox_3$) located at the points $\{{\bf
y}^-_I\}_{1\le I\le\cN}$. If so, the collection of the dislocations is approximately described by the effective energy ${\CW}$, \cite{mal3}:
\begin{eqnarray}
{\CW} = \displaystyle{
\frac{- \mu}{4\pi} \sum\limits_{I,
J} \bigl( {\CU}(\kappa |{\bf y}^+_I-{\bf y}^+_J|)\,+\,
{\CU}(\kappa |{\bf y}^-_I-{\bf y}^-_J|)
\,-\,2\,{\CU}(\kappa |{\bf y}^+_I-{\bf y}^-_J|)
\bigr)}\,, \label{cor:eq15}\\
{\CU}(s)\equiv
\log\bigl(\frac\ga 2 s\bigr)\,+\,K_0(s)\,.
\label{cor:eq151}
\end{eqnarray}
The energy ${\CW}$ (\ref{cor:eq15})
demonstrates that the array of the modified screw dislocations is equivalent to the two-dimensional Coulomb gas of unit charges $\pm 1$ characterized by the two-body potential ${\CU}$ (\ref{cor:eq151}) which is logarithmic at large separation but tends to zero for the charges sufficiently close to each other.

The gauge approach to the statistical physics of the line-like defects (of dislocations and disclinations, in particular) is developed in great detail in \cite{kl12, mult}.
The dislocation core energy is obtained and discussed in \cite{kl12} in order to formulate the lattice model of the defect melting and the corresponding disorder field theory. The dislocation core energy \cite{kl12} (see also \cite{doub1} and \cite{doub2}) depends quadratically on the defect tensor, which is expressed through the disclination density and the  derivatives of first order of the dislocation density.

On the other hand, the core self-energy has been introduced in \cite{mal} in the translationally gauge invariant way by means of the Hilbert--Einstein Lagrangian. The translational gauge Lagrangians in \cite{val, ed1, laz2} are
quadratic in the gauge field strength, i.e., in the dislocation density.
Remind that the geometric theory of defects proposed in \cite{vol} is based on the most general eight-parameter gauge Lagrangian invariant with respect of localized action of three-dimensional Euclidean group. Since the dislocation density is identified as the differential-geometric torsion, \cite{kl12, mult}, the gauge Lagrangians \cite{val, ed1, mal, laz2} may be viewed as parts of the gauge Lagrangian \cite{vol}.

The dislocation core energy in \cite{mal3} originates, by means of linearizations, from the Lagrangian \cite{mal} which in turn can be related to the gauge Lagrangians quadratic in torsion. The action functional proposed in \cite{mal3} for collection
of non-singular dislocations enables the modified screw dislocation to arise as its saddle point, and eventually the effective energy of the modified screw dislocations is given by (\ref{cor:eq15}) and (\ref{cor:eq151}).

The grand-canonical partition function of the Coulomb system described by the energy (\ref{cor:eq15}) takes the form in the dipole phase \cite{ab}:
\begin{equation}
\begin{array}{rcl}
{\bf Z}_{\rm
dip}&=&\displaystyle{\sum \limits_{{\cN}=0}^\infty\,\,
\frac{1}{{\cN}!} \int e^{-2\be{\cN}\La - \be \CW_{\rm
dip}} \prod_{I=1}^{\cN}
\d^2{\bxi}_{I}\d^2{\bta}_{I} }\,,
\\[0.4cm]
\CW_{\rm dip} &\equiv& \displaystyle{\sum\limits_{I}
{w}({\eta}_{I})\,+\,\sum\limits_{I<J}
{w}_{I\! J}}\,,\qquad \beta
w({\eta})\equiv\cK\CU(\kappa \eta)\,,
\end{array}
\label{cor:partf7}
\end{equation}
where $\be=\frac 1T$ is inverse temperature (the Boltzmann constant is unity), ${\cN}$ is the number of
dipoles, $\La$ is the chemical potential per dislocation, and $\cK\equiv\frac{\mu\be}{2\pi}$.
Position of $I^{\rm th}$ dipole ($1\le
I\le\cN$) is given by its center of mass,
${\bxi}_{I}=({\bf y}^+_I+{\bf y}^-_I)/2$,
and momentum, ${\bta}_{I}={\bf y}^+_I-{\bf
y}^-_I$. Averaging over the positions is
replaced by integration over the cylinder's cross-section. The dipole
energy is $w({\eta})$, where
$\eta\equiv|{\bta}|$, while $w_{I\! J}$ is the energy of interaction between $I^{\rm th}$ and $J^{\rm th}$ dipoles:
\begin{equation}
\be w_{I\! J} \,\equiv\,- \cK\,
 (\bta_I, {\boldsymbol\cd}_{{\bxi}_{I}}) (\bta_{J}, {\boldsymbol\cd}_{{\bxi}_{J}})\,  \CU(\kappa|{\bxi}_I-{\bxi}_{J}|)\,,
\label{cor:dip312}
\end{equation}
where the notation ${\boldsymbol\cd}_{{\bxi}}$ implies 2-vector $({\cd}_{{\xi}_{1}}, {\cd}_{{\xi}_2}) \equiv
\bigl(\frac{ \cd}{\cd {{\xi}_1}}, \frac{ \cd}{\cd {{\xi}_2}}\bigr)$ and $(\bta_I, {\boldsymbol\cd}_{{\bxi}_{I}})$ is the scalar product of 2-vectors $\bta_I$ and ${\boldsymbol\cd}_{{\bxi}_{I}}$.

We consider the stress--stress correlation function
$\lav{\si}_{i}({\bf x}_1)\,{\si}_{j}({\bf
x}_2)\rav$ in the dipole representation of the Coulomb gas:
\begin{equation}
\lav{\si}_{i}({\bf x}_1) {\si}_{j}({\bf
x}_2)\rav = \displaystyle{\frac{-\mu}{2\pi \be} \cd_{({\bf x}_1)_i}\cd_{({\bf x}_2)_j} \CU(\kappa|{\bf x}_1-{\bf x}_2|)}+\displaystyle{\frac1{{\bf
Z}_{\rm dip}}
\sum\limits_{{\rm n} \& {\rm
p}}
{\si}_{i}({\bf x}_1) {\si}_{j}({\bf
x}_2)\,e^{-\be\CW_{\rm
dip}}}\,, \label{cor:dip1}
\end{equation}
where $\sum_{{\rm n} \& {\rm p}}$ is
summation over number of dipoles and their positions, $\CW_{\rm
dip}$,
${\bf Z}_{\rm dip}$ are given by
(\ref{cor:partf7}), and ${\si}_{i} = \mu \epsilon_{i k}\,\cd_{{x}_k} \tilde\CU$ (here $\tilde\CU$ is superposition of the stress potentials of $\cN$ dipoles). The second term on the right-hand side of (\ref{cor:dip1}) is specified:
\begin{equation}
\displaystyle{\frac{\mu^2
}{4\pi^2}\,\epsilon_{i k} \epsilon_{j
l}\,{\cd}_{({\bf x}_1)_k} {\cd}_{({\bf x}_2)_l}
\begin{pmatrix}\displaystyle{\frac1{{\bf Z}_{\rm
dip}}\sum\limits_{{\cN}=0}^\infty\,\,
\frac{{\sf F}_{\cN}({\bf x}_1, {\bf
x}_2)}{{\cN}!}}\end{pmatrix}}\,, \label{cor:genf91}
\end{equation}
where
\begin{eqnarray}
{\sf F}_{\cN}({\bf x}_1, {\bf
x}_2)\,\equiv\,
\int e^{-2 \be\cN\La - \be\sum\limits_I {w}({\eta}_{I}) } \prod\limits_{P<Q} e^{-\be {w}_{P Q}}
\nonumber\\
\times\,\Biggl(\sum \limits_{K, L=1}^{\cN}
 \CU^{+-}_K({\bf x}_1)
 \CU^{+-}_L({\bf x}_2)\Biggr)
 \prod\limits_{I=1}^{\cN}
 \d^2{\bxi}_{I}\d^2{\bta}_{I}
 \label{cor:genf92}
\end{eqnarray}
at ${\cN}\ge 1$, and ${\sf F}_{0}=1$. The integration over the dipole positions in
(\ref{cor:genf92}) is analogous to that in (\ref{cor:partf7}). In addition,
$\CU^{+-}_K({\bf x})\equiv \CU^{+-}_{{\bta}_K}({\bf x})$ in (\ref{cor:genf92})
is the stress potential of K$^{\rm th}$
dipole observed at the point ${\bf x}$, and it is approximated at $|\bta_K|\ll |{\bf x}-{\bxi}_K|$:
\begin{equation} \CU^{+-}_{{\bta}_K}({\bf
x})\,\equiv\, \CU(\kappa|{\bf x}-{\bf
y}^+_K|)\,-\,\CU(\kappa|{\bf x}-{\bf
y}^-_K|)\,\approx\,-\,(\bta_K, {\boldsymbol\cd}_{{\bf
x}})\,\CU(\kappa|{\bf x}-{\bxi}_K|)\,,
\label{cor:dip3}
\end{equation}
where ${\boldsymbol\cd}_{{\bf
x}}=({\cd_{x_1}}, {\cd_{x_2}})$. The dipoles are very compact since the dipole momenta are not too large at small enough temperature: $\l {\bta}^2 \r \ll\kappa^{-2}$ \cite{mal3}. We expand $e^{-\be {w}_{P Q} }$ in
(\ref{cor:genf92}) and evaluate (\ref{cor:genf91}) in leading order with respect to $\l {\bta}^2 \r$ thus neglecting all higher momenta $\l {\bta}^4 \r$, etc..
Therefore, Eq.~(\ref{cor:dip1}) takes in the main approximation the form (see \cite{ab} and refs. therein):
\begin{eqnarray}
\lav{\si}_{i}({\bf x}_1)\,{\si}_{j}({\bf
x}_2) \rav \approx \displaystyle{\frac{-\mu}{2\pi
\be}\, \cd_{({\bf x}_1)_i}\cd_{({\bf
x}_2)_j}
\CU(\kappa|{\bf x}_1-{\bf x}_2|)}
\nonumber \\
+ \displaystyle{\frac{\mu^2
}{4\pi^2}\,\epsilon_{i k} \epsilon_{j
l}\,\cd_{({\bf x}_1)_k} \cd_{({\bf x}_2)_l}\sum_{{\tilde n}=1}^\infty {\cal I}_{\tilde n}({\bf x}_1, {\bf x}_2)}\,.\label{cor:dip21}
\end{eqnarray}
The terms ${\cal I}_{\tilde n} \equiv {\cal I}_{\tilde n} ({\bf x}_1, {\bf x}_2)$ in (\ref{cor:dip21}) are expressed:
\begin{eqnarray}
&& \!{\cal I}_{1} \equiv \int e^{-\be(2\La + {w}({\eta}))}\,\CU^{+-}_{\bta}({\bf
x}_1)\,\CU^{+-}_{\bta}({\bf x}_2)\,
\d^2{\bxi}\d^2{\bta}\,, \label{cor:dip211}
\\
&&\! {\cal I}_{\tilde n} \equiv \int e^{-\be(4\La + {w}_{1} +
{w}_{{\tilde n}})} \CU^{+-}_1({\bf
x}_1)
{w}^{({\tilde n}-2)}_{1, {\tilde n}} \CU^{+-}_{\tilde n}({\bf x}_2)\,
\d^2{\bxi}_{1}\d^2{\bta}_{1}
\d^2{\bxi}_{{\tilde n}}\d^2{\bta}_{{\tilde n}}\,, \quad {\tilde n}>1\,, \label{cor:dip22}
\end{eqnarray}
where $w_I\equiv w (\eta_I)$, and $\CU^{+-}_{\bta}({\bf
x}_{1, 2})$ is given by (\ref{cor:dip3}).
The contribution ${\cal I}_{1}$ (\ref{cor:dip211}) is due to free dipoles.
It is the dipole-dipole coupling which is responsible for the terms ${\cal I}_{\tilde n}$ (\ref{cor:dip22}) corresponding to
the ${\tilde n}$-dipole chain clusters.
The kernel ${w}^{({\tilde n}-2)}_{1, {\tilde n}}$ (\ref{cor:dip22}) at ${\tilde n}>2$ is expressed by the recursive relation:
\begin{equation}
w^{(I)}_{1, I+2} = \int e^{-\be(2\La + {w}_{I+1})}
w^{(I-1)}_{1, I+1}\,w^{(0)}_{I+1, I+2}\,\d^2{\bxi}_{I+1} \d^2{\bta}_{I+1}\,,\quad 1\le I \le {\tilde n}-2\,,
\label{cor:dip311}
\end{equation}
where $w^{(0)}_{J K} \equiv -\be w_{J K}$ with $w_{J K}$ given by (\ref{cor:dip312}). The kernel ${w}^{({\tilde n}-2)}_{1, {\tilde n}}$ contains averaging over positions of $({\tilde n}-2)$ ``intermediate'' dipoles.

\section{Calculation of the correlation function}
\label{sec3}

In order to evaluate the integrals (\ref{cor:dip211}) and (\ref{cor:dip22}) we need the following

\vskip0.3cm
\noindent {\bf Proposition:\,} \textit{Define the differential operator
$({\sf D}_\kappa)_{_l}\,\equiv\, \prod_{n=0}^{l-1} ({\sf D}_\kappa+n)$,
$l\ge 1$, where ${\sf D}_\kappa$ stands for $\frac{-\kappa}{2}
\frac{\rm d}{\rm d\kappa}$, and assume that $({\sf D}_\kappa)_{_0}\equiv 1$}.

\noindent{\textsf{A)}} \textit{With these notations, the formula holds}:
\begin{equation}
\int\!K_0 (\kappa| {\bx}_1-{\bxi}|)\, ({\sf
D}_\kappa)_{_l} \, K_0
(\kappa|{\bxi}-{\bx}_{2}|) {\rm d}^2 {\bxi}\,=\, \frac{2\pi}{\kappa^2(l+1)} ({\sf
D}_\kappa)_{_{l+1}}K_0
(\kappa|{\Dl}{\bx}|)\,,\quad l\ge 0,
\label{cor:eq2333}
\end{equation}
\textit{where ${\Dl}{\bx}\equiv {\bx}_{1}-{\bx}_{2}$ and the integration is over} ${\mathbb R}^2$.

\noindent{\textsf{B)}} \textit{Validity of Eq.}~(\ref{cor:eq2333}) \textit{results in the following integral}:
\begin{equation}
\int\! \bigl({\boldsymbol\cd}_{{\bxi}} \CU(\kappa| {\bx}_1-{\bxi}|), {\boldsymbol\cd}_{{\bxi}} ({\sf
D}_\kappa)_{l} \, K_0
(\kappa|{\bxi}-{\bx}_{2}|)\bigr) \d^2 {\bxi}\,=\, \frac{-2\pi}{l+1} ({\sf
D}_\kappa)_{l+1}K_0
(\kappa|{\Dl}{\bx}|)\,,
\label{cor:eq2332}
\end{equation}
\textit{where ${\boldsymbol\cd}_{{\bxi}}=\bigl( {\cd_{\xi_1}}, {\cd_{\xi_2}}\bigr)$, $({\boldsymbol\cd}_{{\bxi}}\,\,\cdot , {\boldsymbol\cd}_{{\bxi}}\,\,\cdot )$ stands for the scalar product of $2$-vectors, and $\CU$ is defined by} (\ref{cor:eq151}).

\noindent Proof of this Proposition is given in Appendix A.

First, we consider the integral (\ref{cor:dip22}) which is expressed through the kernel ${w}^{({\tilde n}-2)}_{1, {\tilde n}}$. After $I\ge 1$ steps, the recursive relation (\ref{cor:dip311}) leads us to the answer:
\begin{eqnarray}
&&{w}^{(I)}_{1, I+2}\,=\,\cK^{I+1} (-\pi \l {\bta}^2 \r \bar N)^{I} (\bta_1, {\boldsymbol\cd}_{{\bxi}_{1}}) (\bta_{I+2}, {\boldsymbol\cd}_{{\bxi}_{I+2}}) \CU^{(I)}_{1, I+2}\,, \label{cor:eq241}\\
&&\CU^{(I)}_{1, I+2}\,=\,\CU(\kappa|{\bxi}_1 - {\bxi}_{I+2}|)\,+\,
\sum\limits_{l=1}^{I} \frac{1}{l!}({\sf
D}_\kappa)_{l} K_0
(\kappa|{\bxi}_1 - {\bxi}_{I+2}|)\,.
\label{cor:eq242}
\end{eqnarray}
Equation (\ref{cor:eq241}) at $I=0$ is  also correct provided that $\CU^{(0)}_{1, 2}$ is given by (\ref{cor:eq242}) with the sum rejected. As an example, let us derive ${w}^{(I)}_{1, I+2}$ at $I=1$. It is appropriate to express the integration over ${\bta}_2$ with the help of definition of mean square of the dipole momentum $\l {\bta}^2 \r$
\cite{kos1, ab}:
\begin{equation}
\displaystyle{\int e^{-2\be\La - {\cK}
\CU(\kappa\eta)} {\eta}_i}
{\eta}_j\,\d^2{\bta}\,=\,\frac{\dl_{i
j}}{2} \l {\bta}^2 \r \bar N\,,
\label{cor:eq19}
\end{equation}
where $\bar N$ is average dipole
density (note that the dipolar phase
does not exist at the temperature $T >
T_c\equiv \frac{\mu}{8\pi}$ since the integral (\ref{cor:eq19})
diverges at $\cK < 4$). Then we obtain:
\begin{equation}
{w}^{(1)}_{1, 3}\,=\,{\cK}^2 \frac{\l {\bta}^2 \r \bar N}{2}
(\bta_1, {\boldsymbol\cd}_{{\bxi}_{1}}) (\bta_{3}, {\boldsymbol\cd}_{{\bxi}_{3}}) \int ({\boldsymbol\cd}_{{\bxi}_2} \CU(\kappa|{\bxi}_1 - {\bxi}_{2}|), {\boldsymbol\cd}_{{\bxi}_2} \CU(\kappa|{\bxi}_2 - {\bxi}_{3}|) ) \d^2{\bxi}_2 \,.
\label{cor:eq251}
\end{equation}
In order to calculate (\ref{cor:eq251}), we use (\ref{cor:eq2332}) at $l=0$ and take into account another integral which can be obtained by means of the Green's first identity \cite{stew}:
\begin{equation}
\int  ({\boldsymbol\cd}_{{\bxi}_2} \log|{\bxi}_1 - {\bxi}_{2}| , {\boldsymbol\cd}_{{\bxi}_2} \CU(\kappa|{\bxi}_2 - {\bxi}_{3}|))  \d^2{\bxi}_2\,=\,2\pi\bigl( \log\frac\gamma 2 R - \CU(\kappa|{\bxi}_1 - {\bxi}_{3}|)\bigr)\,,
\label{cor:eq26}
\end{equation}
where the integration is over disk of large radius $R$ (see the part {\sf B} in Appendix A).
Since the constant contribution on the right-hand side of (\ref{cor:eq26}) is irrelevant with regard to the differentiations in (\ref{cor:eq251}), we can safely allow for $R\to\infty$ and conclude that ${w}^{(1)}_{1, 3}$ is indeed given by Eq.~(\ref{cor:eq241}) at $I=1$ with $\CU^{(1)}_{1,3}$ of the form:
\begin{equation}
\CU^{(1)}_{1,3} = \CU(\kappa|\Dl{\bxi}|)+ {\sf D}_\kappa K_0
(\kappa|\Dl{\bxi}|)\,,
\label{cor:eq252}
\end{equation}
where $\Dl{\bxi}={\bxi}_1 - {\bxi}_{3}$. Next steps based on (\ref{cor:eq2332}) and (\ref{cor:eq26}) lead us straightforwardly to Eqs.~(\ref{cor:eq241}), (\ref{cor:eq242}). The notation ${w}^{({\tilde n}-2)}_{1, {\tilde n}}$ corresponds to (\ref{cor:eq241}) with $I$ replaced by ${\tilde n}-2\ge 0$. The remaining integrations over ${\bxi}_{1}$, ${\bta}_{1}$, ${\bxi}_{{\tilde n}}$, ${\bta}_{{\tilde n}}$ are also enabled by the Proposition, and we obtain that ${\cal I}_{\tilde n}$ (\ref{cor:dip22}) is given in terms of the two-body potential $\CU^{({\tilde
n})}_{1, 2}$:
\begin{equation}
{\cal I}_{\tilde
n}\,=\,\cK^{\tilde n-1} (-\pi \l {\bta}^2 \r \bar N)^{\tilde n}\,\CU^{({\tilde
n})}_{1, 2}\,,\qquad \CU^{({\tilde
n})}_{1, 2} \equiv
\CU(\kappa|{\Dl}{\bx}|)\,+\,
\sum\limits_{l=1}^{\tilde n} \frac{1}{l!}({\sf
D}_\kappa)_{l} K_0
(\kappa|{\Dl}{\bx}|)\,. \label{cor:eq233}
\end{equation}
It can be verified with the help of (\ref{cor:dip211}) and  (\ref{cor:dip311}) that ${w}^{(1)}_{1, 3}$ is formally written in the form:
\begin{equation}
{w}^{(1)}_{1, 3}\,=\,{\cK}^2
(\bta_1, {\boldsymbol\cd}_{{\bxi}_{1}}) (\bta_{3}, {\boldsymbol\cd}_{{\bxi}_{3}}) {\cal I}_{1}({\bxi}_{1}, {\bxi}_{3})\,.
\label{cor:eq23300}
\end{equation}
Comparison of (\ref{cor:eq23300}) with (\ref{cor:eq241}) shows us that ${\cal I}_{1}$ (\ref{cor:dip211}) is also given by the representation (\ref{cor:eq233}).

Definition
of the operator $({\sf
D}_\kappa)_{l}$ tells us that $\CU^{({\tilde
n})}_{1, 2}$ (\ref{cor:eq233}) can be rearranged at fixed $\tilde n$ as the polynomial in powers of ${\sf D}_\kappa$. Then, ${\cal I}_{\tilde
n}$ takes the form:
\begin{equation}
{\cal I}_{\tilde
n}\,=\,\frac{(-\be\mu d)^{\tilde
n}}{\cK}\Bigl(\CU + \sum_{p=1}^{\tilde
n} a_p (\tilde
n) ({\sf
D}_\kappa)^p K_0 \Bigr)\,,\label{cor:eq281}
\end{equation}
where $\be \mu d \equiv \pi\cK \l {\bta}^2 \r\bar N$ and $d$ is proportional to mean area covered by the dipoles. We directly obtain several coefficients $a_p\equiv a_p(\tilde
n)$:
\begin{equation}
a_1\,=\,\sum_{l=1}^{\tilde
n}\frac 1l\,,\quad a_2\,=\,\sum_{l=2}^{\tilde
n}\frac1l \Bigl(\sum_{k=1}^{l-1}\frac 1k\Bigr)\,, \ldots,\quad a_{\tilde
n-1}\,=\,\frac {\tilde
n+1}{2({\tilde
n-1})!}\,,\quad\,a_{\tilde
n}\,=\,\frac 1{{\tilde
n}!}
\,.
\label{cor:eq282}
\end{equation}
For instance, $a_1=1$ and $a_l=0$, $l> 1$, for ${\tilde n}=1$. More generally,
non-triviality of $a_l({\tilde n})$ requires ${\tilde n}\ge l$. The coefficients $a_{\tilde n-1}({\tilde n})$ and $a_{\tilde n}({\tilde n})$, Eq.~(\ref{cor:eq282}), taken at ${\tilde n}=2$ coincide with $a_1(2)$ and $a_2(2)$, respectively. Besides, $a_{\tilde n-1}({\tilde n})$ (\ref{cor:eq282}) at ${\tilde n}=3$ is equal to $a_2(3)$.

It is appropriate to rearrange the series ${\cal I} \equiv \sum_{{\tilde
n}=1}^\infty {\cal I}_{\tilde
n}$ in order to represent it as an expansion in $({\sf D}_\kappa)^p K_0$. For instance, using re-summation we
obtain the series at $\be\mu d<1$:
\begin{eqnarray}
&&\sum_{{\tilde n}=1}^\infty (-\be\mu d)^{\tilde n} a_1({\tilde n})\,=\,\sum_{{l}=0}^\infty (-\be\mu d)^{l} \sum_{{\tilde n}=1}^\infty \frac{(-\be\mu d)^{\tilde n}}{\tilde n}\,=\,\frac{-\log(1+\be\mu d)}{1+\be\mu d} \,,\label{cor:eq2831}\\
&&\sum_{{\tilde n}=2}^\infty (-\be\mu d)^{\tilde n} a_2({\tilde n})
\,=\,\sum_{{l}=0}^\infty (-\be\mu d)^{l} \sum_{{\tilde n}=2}^\infty \frac{(-\be\mu d)^{\tilde n}}{\tilde n}\Bigl(\sum_{k=1}^{{\tilde n}-1}\frac1k\Bigr)\,=\,\frac{\log^2(1+\be\mu d)}{2 (1+\be\mu d)}\,,
\label{cor:eq2832}
\end{eqnarray}
where \cite{brych} (vol.1) is helpful for obtaining (\ref{cor:eq2832}).
Therefore ${\cal I}$ takes the form:
\begin{equation}
{\cal I}\,=\,\frac{-1}{\cK}\Bigl[\CU \frac{\be\mu d}{1+\be\mu d} + {\sf
D}_\kappa K_0 \frac{\log(1+\be\mu d)}{1+\be\mu d} - ({\sf
D}_\kappa)^2 K_0 \frac{\log^2(1+\be\mu d)}{2(1+\be\mu d)} + \ldots \Bigr]\,.
\label{cor:eq29}
\end{equation}

Consider the dependence of ${\cal I}$ (\ref{cor:eq29}) on
$\kappa |{\Dl}{\bx}|$. The contribution $\CU$ is long-ranged, while its behavior at $\kappa |{\Dl}{\bx}| \to 0$ is that of the modified Bessel function $K_0$ which ensures the cancellation of the logarithmic singularity. The contributions $({\sf D}_\kappa)^p K_0$, $p\ge 1$, are exponentially decaying at large $\kappa |{\Dl}{\bx}|$. They all vanish at $\kappa |{\Dl}{\bx}|\to 0$ except the term ${\sf
D}_\kappa K_0= \frac{\kappa |{\Dl}{\bx}|}{2}K_1
(\kappa |{\Dl}{\bx}|)$ which is constant at $\kappa |{\Dl}{\bx}|\to 0$.
The third term in (\ref{cor:eq29}) vanishes at $\kappa|\Dl|\to 0$ faster than $\CU(\kappa |\Dl|)$ since
$({\sf D}_\kappa)^2 K_0 (\kappa |\Dl|)=
\frac{(\kappa |\Dl|)^2}{4} K_0 (\kappa |\Dl|)$. Furthermore, $\log (1+\be\mu d)< 1$ and we omit the third term (as well as all the subsequent ones) as negligible correction thus adopting for (\ref{cor:genf91}) the approximate expression:
\begin{equation}
\begin{array}{rcl}
&&
\lav{\si}_{i} ({\bf x}_1)\,{\si}_{j}({\bf x}_2)\rav\,\approx\,\displaystyle{
\frac{-\mu}{2\pi \be}}\,\Bigl(\cd_{({\bf x}_1)_i}\cd_{({\bf x}_2)_j} \CU(\kappa |{\Dl}{\bx}|)\,+\Bigr.\, \\[0.3cm]
&& +\,\Bigl. \displaystyle{
\epsilon_{i k}
\epsilon_{j l}\,\cd_{({\bf x}_1)_k}
\cd_{({\bf x}_2)_l} \frac{ \be\mu d\,\CU(\kappa |{\Dl}{\bx}|)+\log(1+\be\mu d)\,{\sf
D}_\kappa K_0 (\kappa |{\Dl}{\bx}|)}{1+\be\mu d}}\Bigr) \,.\label{cor:eq27}
\end{array}
\end{equation}

When the short-ranged contributions into the ``intermediate'' dipole-dipole correlations are neglected, the kernel $w^{({\tilde n}-2)}_{1, {\tilde n}}$ is given by Eq.~(\ref{cor:eq241}) although being expressed in terms of the two-body potential $\CU^{({\tilde n}-2)}_{1, {\tilde n}} \approx \log|{\bxi}_1- {\bxi}_{\tilde n}|$ (see (\ref{cor:eq242})). If so, evaluating ${\cal I}_{\tilde
n}$ according to (\ref{cor:dip22}) we obtain ${\cal I}_{\tilde
n}\,=\,\cK^{\tilde n-1} (-\pi \l {\bta}^2 \r \bar N)^{\tilde n}\,\CU^{(1)}_{1, 2}$
instead of ${\cal I}_{\tilde
n}$ (\ref{cor:eq233}).
In this case, the correlation function (\ref{cor:dip21}) takes the form similar to (\ref{cor:eq27}) although with $\be\mu d$ present instead of $\log (1+\be\mu d)$. In other words, the presence of $\log (1+\be\mu d)$ in (\ref{cor:eq27}) just witnesses that the correlations due to the cores are taken into account for the dipole-dipole couplings. The latter is in agreement with the fact that $\log (1+\be\mu d)$ in (\ref{cor:eq27}) is replaced by $\be\mu d$ when the density is not too large.

\section{The shear modulus renormalization}
\label{sec4}

The stress--stress correlation function (\ref{cor:dip1}) is used for definition of the renormalized shear modulus $\mu_{\rm ren}$, \cite{nel12, rab}:
\begin{equation}
\label{cor:ren1} \displaystyle{\frac{1}{
\mu_{\rm ren}}\,\equiv\,\frac{\be}{\mu^2
\cS}\,\sum\limits_{i, k}
\,\int}\lav{\si}_{i}({\bf
x}_1)\,{\si}_{k}({\bf x
}_2)\rav\,\d^2{\bx}_1 \d^2{\bx}_2\,,
\end{equation}
where $\cS$ is the cross-section area. Substituting the expression
(\ref{cor:eq27}) into Eq.~(\ref{cor:ren1}) yields:
\begin{equation}
\displaystyle{\frac{\mu}{\mu_{\rm
ren}}\,=\,\frac{(1+2 \be\mu d)\,{\cC}_1(\kappa R)-\log(1+\be\mu d)\,{\cC}_{\sf D}(\kappa R)}{1+\be\mu d}
}\,, \label{cor:ren21}
\end{equation}
where the functions ${\cC}_{1}$ and ${\cC}_{\sf D}$ are given by the modified Bessel functions:
\begin{equation}
\begin{array}{rcl}
&&{\cC}_1(z)= 1-2K_1(z) I_1(z)\,,\\[0.3cm]
&&{\cC}_{\sf D}(z)=- {\sf D}_z
{\cC}_1(z)\,=\,-1+z I_1(z) (K_0(z)+K_2(z))\,.
\end{array}
\label{cor:ren22}
\end{equation}
The functions (\ref{cor:ren22}) are responsible for the dependence of $\mu_{\rm ren}$ (\ref{cor:ren21}) on the ratio $R/\kappa^{\1}$ of the cylinder's cross-section radius to that of the dislocation core. The functions ${\cC}_{1}$ and ${\cC}_{\sf D}$ are positive and less than unity (Figure~1), their behavior at large $z$ is: ${\cC}_1(z) =
1 - \frac{1}{z} +\frac{3}{8
z^3}-\dots$ and ${\cC}_{\sf D}(z) =\frac{1}{2z} -\frac{9}{16 z^3} +\dots$. Both
${\cC}_1$ and ${\cC}_{\sf D}$ are
${\cal O} (z^2\log z)$ at $z\to 0$.
\begin{figure}[h]
\centering
\includegraphics[scale=0.40]{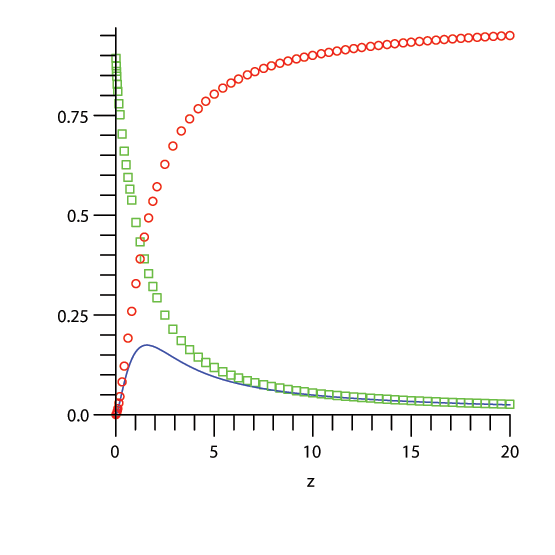}
\caption{The functions ${\cC}_1(z)$ ($\circ$), ${\cC}_{\sf D}(z)$, and $\frac{{\cC}_{\sf D}(z)}{{\cC}_1(z)}$ (${\square}$).}
\end{figure}

The limit $\kappa R\to\infty$ reduces Eq.~(\ref{cor:ren21}) to the renormalization law for conventional dislocations
described by the stress ${\si}^{\rm b}_{i}({\bf x})$. It can be realized that the correlation function $\lav{\si}^{\rm b}_{i}({\bf
x}_1)\,{\si}^{\rm b}_{k}({\bf x
}_2)\rav$ coincides with right-hand side of Eq.~(\ref{cor:eq27}) taken at $\kappa|\Dl{\bf x}|\gg 1$, and we get:
\begin{equation}
\sum\limits_{k} \lav{\si}^{\rm b}_{k}({\bf
x}_1)\,{\si}^{\rm b}_{k}({\bf x
}_2)\rav \,=\,\frac{\mu (1+2 \be\mu d)}{\be(1+\be\mu d)}\,{\stackrel{(2)}
\dl}({\bx}_1-{\bx}_2)\,.
\label{cor:ren221}
\end{equation}
Substituting (\ref{cor:ren221}) into (\ref{cor:ren1}) we obtain (``non-diagonal'' correlators at $i\ne k$ are irrelevant with respect to the integrations in (\ref{cor:ren1})):
\begin{equation}
\displaystyle{\frac12\,<\,\frac{\tilde
\mu_{\rm
ren}}{\mu}\,=\,\frac{1+\be\mu d}{1+ 2 \be\mu d}\,<\,1}\,,
\label{cor:ren222}
\end{equation}
where $\tilde \mu_{\rm ren}$ is the renormalized shear modulus in the conventional case. It is clear that (\ref{cor:ren222}) arises from (\ref{cor:ren21}) at $\kappa R\to\infty$. The law (\ref{cor:ren222}) agrees with the renormalization of the shear modulus derived in \cite{rab} although for the dislocation loops in three-dimensional solid.

Let us specify a range of $\kappa R$. For instance, validity of $0 < {\cC}_{1} < 1$ implies that ${\mu_{\rm ren}}/{\mu}$ is characterized by the double-sided inequality analogous to  (\ref{cor:ren222}):
\begin{equation}
\frac12\,<\,
\frac{1}{2{\cC}_1}\,<\,
\frac{\mu_{\rm ren}}{\mu}\,
<\,1\,.
\label{cor:ren5}
\end{equation}
Inequality (\ref{cor:ren5}) is not fulfilled immediately. Indeed, its self-consistency requires ${\cC}_{1}>\frac 12$ which takes place at $\kappa R\,{\stackrel{>}{_\sim}}\,1.78$.
Furthermore, since external diameter of the nanotubes ranges from nanometers to tens of
nanometers, let us assume that the cylinder's radius $R$ respects $R \,{\stackrel{<}{_\sim}}\, 10\,a$, where $a$ is lattice spacing. In turn, it is known that the inverse $\frac{1}\kappa$
is estimated either as $\frac1\kappa\simeq 0.25a$, according to \cite{g1}, or as
$\frac1\kappa\simeq 0.4a$, according to
\cite{laz2}. Therefore, the upper bound for $\kappa R$ is given either by $\kappa R\,{\stackrel{<}{_\sim}}\,40$ or
by $\kappa R
\,{\stackrel{<}{_\sim}}\,25$, respectively.

Being considered at small $\be\mu d$, the inequality on the right-hand side of (\ref{cor:ren5}) is reformulated:
\begin{equation}
\frac{1+\be \mu d}{(1+2 \be \mu d){\cC}_1(z) - \be \mu d
{\cC}_{{\sf D}}(z)}\,<\,1
\,\Longleftrightarrow\,f_1(z)\,\equiv\,
\frac{1-{\cC}_1(z)}{2\,{\cC}_1(z)-{\cC}_{{\sf D}}(z)-1}
\,<\,\be \mu d\,. \label{cor:ren6}
\end{equation}
The equivalence in (\ref{cor:ren6}) occurs together with positivity and monotonic decreasing of $f_1(z)$, for instance, at $z>2.3$. However, Figure~2 demonstrates us that $f_1(z)<\be\mu d$ only for large enough $z$. On the contrary, when $\be\mu d$ is close to unity, we use $\log(1+\be\mu d)\approx \log 2 +\frac{\be\mu d-1}2$ to re-express right-hand side of (\ref{cor:ren5}):
\begin{equation}
f_2(z)\,\equiv\,
\frac{1- {\cC}_1(z)+(\log 2-\frac12)\,{\cC}_{{\sf D}}(z)}{2\,{\cC}_1(z)-\frac12\, {\cC}_{{\sf D}}(z)-1}
\,<\,\be\mu d\,. \label{cor:ren61}
\end{equation}
Figure~2 demonstrates that $f_2(z)$ is positive, monotonically decreases and almost coincides with $f_1(z)$ at $z\,{\stackrel{>}{_\sim}}5.0$. Therefore, the inequality in
right-hand side of (\ref{cor:ren5}) holds for a more wide range of $z$ in the case of $\be\mu d$ close to unity than in the case of
$\be\mu d$ close to zero.
\begin{figure}[h]
\centering
\includegraphics[scale=0.40]{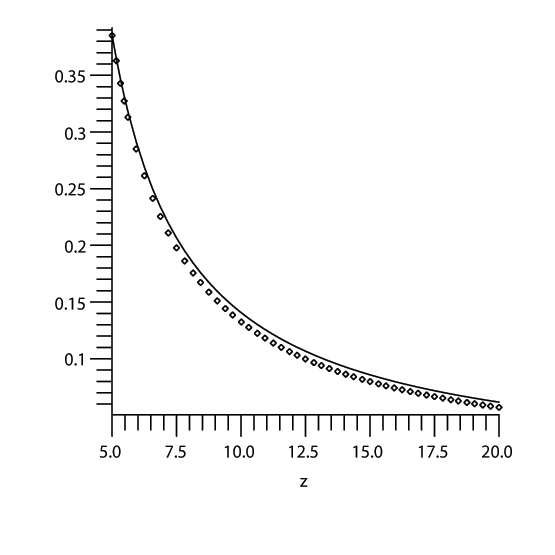}
\caption{The functions $f_1(z)$ ($\diamond$) and $f_2(z)$ at $z=\kappa R>5.0$.}
\end{figure}
Besides,
Eq.~(\ref{cor:ren21}) shows us
that ${\mu_{\rm ren}}/{\mu}$ tends to
$\frac{1}{2{\cC}_{1}}$ provided
that $\be \mu d$ is sufficiently large.

Furthermore, we plot $\mu_{\rm ren}/\mu$ (\ref{cor:ren21}) (for $\kappa R=20.0$ and $\kappa R=40.0$) and $\tilde\mu_{\rm ren}/\mu$ (\ref{cor:ren222}) in Figure~3 as functions of $\be \mu d$.
\begin{figure}[h]
\centering
\includegraphics[scale=0.50]{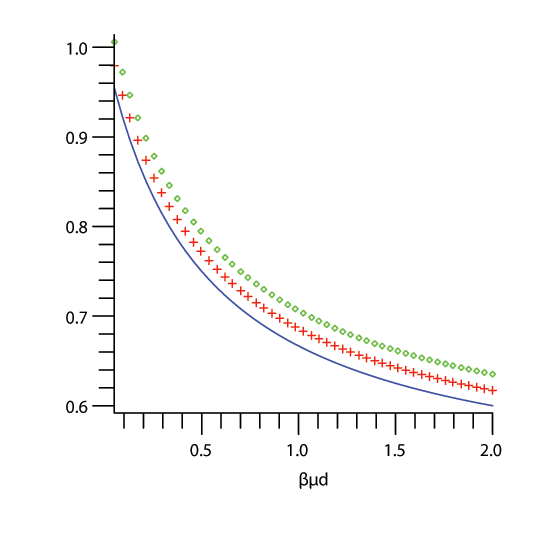}
\caption{Dependence of $\mu_{\rm ren}/\mu$ ($\kappa R=20.0$ ($\diamond$) and $\kappa R=40.0$ ($+$)) and  $\tilde\mu_{\rm ren}/\mu$ on $\be \mu d$.}
\end{figure}
Additionally, the dependence of $\mu_{\rm ren}/\mu$ on $z=\kappa R$ is plotted in Figure~4 for $\be\mu d=0.2,\, 0.5,\, 0.9$.
\begin{figure}[h]
\centering
\includegraphics[scale=0.50]{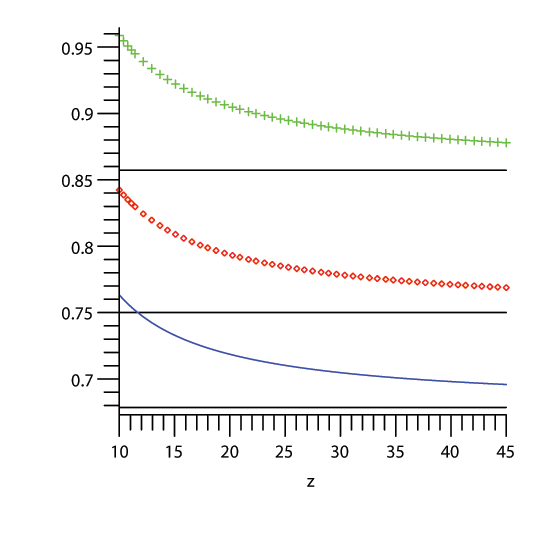}
\caption{Dependence of $\mu_{\rm ren}/\mu$ on $z=\kappa R$ at $\be\mu d=0.2$ ($+$), $\be\mu d=0.5$ ($\diamond$), $\be\mu d=0.9$.}
\end{figure}
The picture shows that at each value of $\be\mu d$ the curves for $\mu_{\rm ren}/\mu$ tend asymptotically to the horizontal lines
corresponding to the constant values of $\tilde\mu_{\rm ren}/\mu$  (\ref{cor:ren222}).

Equation (\ref{cor:ren21}) enables us to
express the renormalized shear modulus as the function of the absolute temperature,
$\mu_{\rm {ren}}= \mu_{\rm {ren}}(T)$. The melting temperature $T_c$
is given by $\mu\be_c=8\pi$ (see (\ref{cor:eq19})), and we obtain
$\mu_{\rm {ren}}(T)$ in a narrow vicinity of $T_c$ at
$T<T_c$:
\begin{equation}
\begin{array}{r}
\displaystyle{\frac{\mu_{\rm
{ren}}(T)}{\mu_{\rm {ren}}(T^-_c)} \approx 1
+ \Bigl(\frac{T}{T_c}-1 \Bigr) h(\kappa
R, 8 \pi d)}\,, \quad h(z, w)\equiv\frac{w\, g(z, w)}{(1+w)(1+2 w)}\,, \\ [0.4cm]
\displaystyle{g(z, w)\equiv
\frac{1-{\cC}^*(z)(1-\log(1+w))}{1- {\cC}^*(z)\frac{\log(1+w)}{1+2w}}}\,,\qquad
{\cC}^*(z)\equiv
\frac{{\cC}_{{\sf D}}(z)}{{\cC}_{1}(z)} \,,
\end{array}
\label{cor:ren7}
\end{equation}
where the representation for $h(z, w)$ is valid at $z>\epsilon_{\rm o}\equiv 0.001$, when ${\cC}^*(z)$ is smaller than unity and the limiting value ${\cC}^*_{\rm o}\equiv\lim_{z\searrow \epsilon_{\rm o}}{\cC}^*(z)$ satisfies $0.75<{\cC}^*_{\rm o}<0.95$ (Figure~1).
In the limit $z \to\infty$, ${\cC}^*(z)$ tends to zero and the law (\ref{cor:ren7}) corresponds to singular dislocations. Furthermore, $\mu_{\rm
{ren}}(T^-_c)$ in (\ref{cor:ren7}) is the limiting value of
$\mu_{\rm {ren}}(T)$ at $T \nearrow T_c$ ($\mu_{\rm {ren}}$ is zero above
the melting). Comparison of $\mu_{\rm {ren}}(T^-_c)$ with
${\tilde\mu}_{\rm {ren}}(T^-_c)$
is enabled, according to (\ref{cor:ren21}), in the form:
\begin{equation}
\frac{\mu_{\rm
{ren}}(T^-_c)}{T_c}\,=\,\frac{8 \pi(1+8 \pi
d)}{(1+16 \pi
d) {\cC}_1 - \log(1+8 \pi  d)
{\cC}_{\sf D}}\,>\,\frac{{\tilde\mu}_{
\rm {ren}}(T^-_c)}{T_c}\,=\,\frac{8 \pi
(1+8 \pi d)}{1 + 16 \pi  d }\,.
\label{cor:ren8}
\end{equation}
Equation (\ref{cor:ren8})
demonstrates that ${\mu_{\rm
{ren}}(T^-_c)}/{T_c}$ ceases to be a multiple of $\pi$ at $d\ll 1$ (or, formally, $d\gg
1$), as it
happens for ${{\tilde\mu}_{ \rm
{ren}}(T^-_c)}/{T_c}$ in the same limits.
In the limiting cases, the factor
$\frac{1}{{\cC}_1}$ is mainly responsible for the dependence of ${\mu_{\rm
{ren}}(T^-_c)}/{T_c}$ on $\kappa R$.

The dependence on the size-parameter $\kappa R$ in Eqs.~(\ref{cor:ren7}) and
(\ref{cor:ren8})  displays the effect of the non-conventional dislocation solution on the shear
modulus near the melting transition.
Recall that properly rescaled Young modulus tends to $16\pi$ at $T\to T_c^-$ according to the theory developed in \cite{holz, nel1, nel12, nel13} (this universality is also discussed in \cite{kl12}). An experimental confirmation of this
fact for two-dimensional colloidal crystals has been reported in \cite{grun}. In its turn, the present approach demonstrates that the limiting value of the renormalized shear modulus  deviates from a multiple of $\pi$ due to the singularityless character of the dislocations:  \begin{equation}
\frac{\mu_{\rm
{ren}}(T^-_c)}{{T_c}}\approx \frac {8\pi}{{\cC}_1(\kappa R)}
\begin{CD}@>> \kappa R\gg 1>\end{CD}\,8\pi \,, \qquad d\ll 1
\label{cor:ren100}
\end{equation}
(note that $1<\frac{1}{{\cC}_1}<2$ under our restrictions).

\section{Discussion}

We considered the behavior of the
shear modulus caused by proliferation of dipoles of non-singular screw dislocations in elastic cylinder. The dislocations are non-singular since their stresses are smoothed out within the finite-sized cores. The stress--stress correlation function is calculated in the dipole representation of the Coulomb gas with smoothed-out coupling. The shear modulus $\mu_{\rm ren}$ is obtained for the case of interacting dipoles. The resulting thermodynamical expressions, Eqs.~(\ref{cor:ren21}), (\ref{cor:ren7}) and (\ref{cor:ren8}), acquire the dependence on the ratio of the cylinder radius $R$ to the core scale $\frac1\kappa$.
The correlations between the cores are responsible for the fact that $\mu_{\rm ren}/\mu$ decays to its limiting value at growing $d$ ($d$ is proportional to mean area covered by the dipoles) less fast than $\tilde\mu_{\rm ren}/\mu$ does in the case of singular defects (Figure~3). The numerical restrictions ensuring the validity of
(\ref{cor:ren21}), (\ref{cor:ren7}) and
(\ref{cor:ren8}) do not contradict to realistic characteristics of
nanotubes (nanowires). Plots in Figure~4 show us that the dependence of $\mu_{\rm ren}/\mu$ on $\kappa R$ is rather fine (in vicinity of the melting transition as well).

With regard to the interesting
results of \cite{grun} based on the colloidal crystals, it is hopeful
that experimental nanophysics could
provide us opportunities of verification of the relations obtained (of Eq.~(\ref{cor:ren100}), for instance). The colloidal crystals have also been mentioned in \cite{rab} among other candidates for observing the effects due to the elastic constants renormalization. Development of the present approach is attractive in the case of nonsingular edge dislocations as far as the physics of multi-layer nanotubes and wrapped crystals is concerned \cite{dkl, graph1}.
\section*{Acknowledgement}

I am grateful to
N.~M.~Bogoliubov for discussions. The research described
has been supported in part by RFBR
(No.~13-01-00336) and by the Russian
Academy of Sciences program ``Mathematical
Methods in Non-Linear Dynamics''.

\section*{APPENDIX A. Proof of Eqs.~(\ref{cor:eq2333}), (\ref{cor:eq2332})}

{\textsf{A)}} First, we obtain the integral (\ref{cor:eq2333}) at $l=0$:
\begin{align*}
&\int\! K_0 (\kappa| {\bx}_1-{\bxi}|) K_0
(\kappa|{\bxi}-{\bx}_{2}|) {\rm d}^2 {\bxi} =
\tag{{\rm A}.1}
\\
&=\,\frac{2}{\kappa^2} \int_0^\infty{\rm d} s s K_0(s)\int_0^{\pi} {\rm d}\varphi K_0\bigl({\sqrt{s^2+a^2-2 a s \cos\varphi } }\bigr)\,=
\tag{{\rm A}.2}
\\
&=\,\frac{\pi}{\kappa}\,|{\Dl}{\bx}| K_1 (\kappa|{\Dl}{\bx}|) = \frac{2\pi}{\kappa^2}\,{\sf
D}_\kappa K_0
(\kappa|{\Dl}{\bx}|)\,,
\tag{{\rm A}.3}
\end{align*}
where $a\equiv \kappa |{\Dl}{\bx}|$ in  (A.2). The resulting expression (A.3) arises as follows: first we integrate in (A.2) over $\varphi$ and then over $s$ using in both cases \cite{brych} (vol.~2). Equation (\ref{cor:eq2333}) can analogously be verified at $l=1$.

A formal way to evaluate left-hand side of (\ref{cor:eq2333}) is provided in \textsl{Appendix~B}, and it can practically be performed for several first values of $l$. Moreover, these particular examples just allow us to postulate Eq.~(\ref{cor:eq2333}) at arbitrary $l$.
Once Eq.~(\ref{cor:eq2333}) has been guessed, its proof is given by complete induction with Eqs.~(A.1)--(A.3) considered as the base case of induction.

The integrals (\ref{cor:eq2333}) and (A.1) express the convolution of two functions. For sake of shortness, we use the convolution notation so that (\ref{cor:eq2333}) acquires the operator form:
\begin{equation}
K_0 \ast ({\sf
D}_\kappa)_{_l} \, K_0
\,=\, \frac{2\pi}{\kappa^2(l+1)} ({\sf
D}_\kappa)_{_{l+1}}K_0\,.
\nonumber
\tag{{\rm A}.4}
\end{equation}
Let us assume that Eq.~(A.4) is respected at some fixed $l$. The inductive step requires to prove that the relation arising from (A.4) after the replacement $l\mapsto l+1$  also holds true due to the validity of (A.3) and (A.4).

In order to proceed, we convolve both sides of (A.4) with $K_0$. Using associativity of convolution and taking into account the basis of induction (A.3), we obtain:
\begin{equation}
\frac{2\pi}{\kappa^2}({\sf
D}_{\kappa}K_0\ast ({\sf
D}_\kappa)_{_{l}} K_0)\,=\, \frac{2\pi}{\kappa^2(l+1)}(K_0 \ast ({\sf D}_\kappa)_{_{l+1}} K_0)\,.
\nonumber
\tag{{\rm A}.5}
\end{equation}
We transform the operator on the left-hand side of (A.5):
\begin{align*}
{\sf
D}_{\kappa}K_0\ast ({\sf
D}_\kappa)_{_{l}} K_0 \,=
{\sf
D}_{\kappa}\Bigl( \frac{2\pi}{\kappa^2(l+1)}({\sf D}_\kappa)_{_{l+1}} K_0\Bigr) -
K_0\ast {\sf
D}_{\kappa} ({\sf D}_\kappa)_{_{l}} K_0
\tag{{\rm A}.6}
 \\
\,=\,\frac{2\pi}{\kappa^2(l+1)}({\sf
D}_{\kappa}+1)({\sf D}_\kappa)_{_{l+1}} K_0 -
K_0\ast {\sf
D}_{\kappa} ({\sf D}_\kappa)_{_{l}} K_0\,.
\tag{{\rm A}.7}
\end{align*}
We used (A.4) in the row
(A.6), while the row (A.7) occurs due to ${\sf D}_{\kappa}(\frac1{\kappa^{2}}) = \frac1{\kappa^{2}}$. Furthermore, we take into account that $({\sf D}_\kappa)_{_{l+1}} = ({\sf D}_\kappa+l)({\sf D}_\kappa)_{_{l}}$ on the right-hand side of (A.5). As a result, Eq.~(A.5) takes the form:
\begin{equation}
\frac{2\pi}{\kappa^2}({\sf
D}_{\kappa}+1)({\sf
D}_\kappa)_{_{l+1}} K_0\,=\, K_0 \ast
\bigl((l+1){\sf D}_\kappa + {\sf D}_\kappa+l\bigr)
({\sf D}_\kappa)_{_{l}} K_0\,.
\nonumber
\tag{{\rm A}.8}
\end{equation}
As the next step, we use the operator identity
\begin{equation}
(l+1){\sf D}_\kappa + {\sf D}_\kappa+l = (l+2)({\sf D}_\kappa +l)-l(l+1)\,,
\nonumber
\end{equation}
and obtain from (A.8):
\begin{equation}
\begin{array}{l}
\displaystyle{
\frac{2\pi}{\kappa^2}({\sf
D}_{\kappa}+1)({\sf
D}_\kappa)_{_{l+1}} K_0\,=\,}\\[0.2cm]
=\,\displaystyle{
(l+2)(K_0\ast ({\sf
D}_\kappa)_{_{l+1}} K_0) -
\frac{2\pi}{\kappa^2}\,l ({\sf
D}_\kappa)_{_{l+1}} K_0}
\end{array}
\nonumber
\tag{{\rm A}.9}
\end{equation}
(Eq.~(A.4) is accounted for once again when obtaining the second term on the right-hand side of (A.9)). Equation (A.9) is just equivalent to (A.4) with $l$ replaced by $l+1$.~$\Box$

\noindent {\textsf{B)}\,} Let us assume that the integration domain in Eq.~(\ref{cor:eq2332}) is given by disk
of large enough radius $R$ with two disks of small radii $\varepsilon$ centered at ${\bx}_{1}$ and ${\bx}_{2}$ cut out.
The integral (\ref{cor:eq2332}) is evaluated by the limiting trick: we replace $\CU(\kappa| {\bx}_1-{\bxi}|)$ by $\CU(\rho| {\bx}_1-{\bxi}|)$ and drop the operator $({\sf D}_\kappa)_{l}$ out the integration symbol. We use Green's first identity \cite{stew} for thus arising integral over the multi-connected domain. Tending $R\to\infty$, $\varepsilon\to 0$  afterwards, we again place $({\sf D}_\kappa)_{l}$ under the integration symbol and replace $\rho\to\kappa$. Eventually, taking into account (\ref{cor:eq2333}), we obtain (\ref{cor:eq2332}).~$\Box$

\section*{APPENDIX B. The generating function of Eq.~(\ref{cor:eq2333})}

Left-hand side of Eq.~(\ref{cor:eq2333}) is also handled at $l\ge 1$ by means of the limiting trick. We replace $({\sf
D}_\kappa)_{l} K_0
(\kappa|{\bxi}-{\bx}_{2}|)$ by $({\sf
D}_\rho)_{l} K_0
(\rho|{\bxi}-{\bx}_{2}|)$ and drop the operator $({\sf D}_\rho)_{l}$ out the integration symbol.
Thus arising integral generalizes the integral (A.1) and admits the series representation:
\begin{align*}
&\int K_0 (\kappa| {\bx}_1-{\bxi}|) K_0
(\rho|{\bxi}-{\bx}_{2}|) {\rm d}^2 {\bxi}\,=
\tag{{\rm B}.1} \\
&=\,\frac{2\pi( K_0(a)- K_0(b))}{\rho^2-\kappa^2}
\tag{{\rm B}.2}
 \\
&\,=\frac{2\pi}{\kappa+\rho}\Biggl[ \frac2\kappa\,{\sf
D}_\kappa K_0(a)+\sum_{p=1}^\infty \frac{2(\kappa-\rho)^p}{(p+1) \kappa^{p+1}}\,{\sf D}_\kappa \prod_{l=1}^p\bigl(1+\frac{2}{l}\,{\sf
D}_\kappa\bigr)K_0(a)\Biggr]\,,
\tag{{\rm B}.3}
\end{align*}
where $a\equiv\kappa |{\Dl}{\bx}|$,
$b\equiv\rho |{\Dl}{\bx}|$. Acting by $({\sf
D}_\rho)_{l}$ on (B.3) and allowing $\rho\to\kappa$ afterwards, one should derive right-hand side of (\ref{cor:eq2333}). In other words, the integral (B.1) is expected to be the generating function of Eq.~(\ref{cor:eq2333}).

Calculation of the integral (B.1) is analogous to that of the integral (A.1), and the answer takes the form (B.2) (see \cite{brych}, vol.~2). Expanding $K_0(b)$ in (B.2) in powers of $(\kappa-\rho)$, we obtain the series representation:
\begin{equation}
\frac{2\pi}{\kappa+\rho}\Biggl[- |{\Dl}{\bx}| K^\prime_0(a)+\sum_{p=1}^\infty
\frac{(- |{\Dl}{\bx}|)^{p+1}(\kappa-\rho)^p}{(p+1)!
} K^{(p+1)}_0(a) \Biggr]\,.
\nonumber
\tag{{\rm B}.4}
\end{equation}
It is straightforward to check that
the ``constant'' terms inside the brackets both in (B.3) and (B.4)
coincide. Induction allows us to prove the coincidence of the coefficients of the series inside the brackets and thus to confirm the series representation (B.3).

The relation expressing the coincidence can be written as
\begin{equation}
a^{p+1} \frac{{\d}^{p+1}}{\d a^{p+1}} K_0(a) = \kappa \frac{{\d}}{\d \kappa}
\prod_{l=1}^p \Bigl(\kappa \frac{{\d}}{\d \kappa} - l\Bigr) K_0(a)\,.
\nonumber
\tag{{\rm B}.5}
\end{equation}
Assuming that Eq.~(B.5) is valid at some fixed $p$, we have to prove its validity after the replacement $p\mapsto p+1$. The identity
$$
a^{p+2} \frac{{\d}^{p+2}}{\d a^{p+2}} =
\Bigl(
a \frac{{\d}}{\d a} - (p+1)
\Bigr)\,
a^{p+1} \frac{{\d}^{p+1}}{\d a^{p+1}}
$$
just allows for the required proof. Thus, Eq.~(B.3) indeed follows from (B.2).

As an illustration, we put $l=2$ and apply $\lim_{\rho\to\kappa} ({\sf
D}_\rho)_{_2}$ to (B.3). Using the limiting expressions
\begin{equation}
\lim_{\rho\to\kappa} ({\sf
D}_\rho)_{_2}\frac{1}{\kappa+\rho} = \frac{1}{8\kappa}\,,\qquad
\lim_{\rho\to\kappa} ({\sf
D}_\rho)_{_2}\frac{(\kappa-\rho)^p}{\kappa+\rho} = \left [\,\,\begin{matrix} \frac14, \quad p=1\,,\\ \frac\kappa 4, \quad p=2\,,\\ 0, \quad p>2\,, \end{matrix}\right.
\nonumber
\end{equation}
we obtain:
\begin{align*}
&\int K_0 (\kappa| {\bx}_1-{\bxi}|) ({\sf
D}_\kappa)_{_2} K_0
(\kappa|{\bxi}-{\bx}_{2}|) {\rm d}^2 {\bxi} = \\
& = \frac{2\pi}{4\kappa^2} \Bigl[ {\sf
D}_\kappa K_0(a)+ {\sf
D}_\kappa(1+2{\sf
D}_\kappa) K_0(a)+\frac23\, {\sf
D}_\kappa(1+2{\sf
D}_\kappa)(1+{\sf
D}_\kappa)K_0(a)\Bigr] = \\
&= \frac{2\pi}{3\kappa^2} {\sf
D}_\kappa (1+{\sf
D}_\kappa) (2+ {\sf
D}_\kappa) K_0(a)\,.
\tag{{\rm B}.6}
\end{align*}
Equation (B.6) agrees with
(\ref{cor:eq2333}) at $l=2$. It can directly be verified for several $l> 2$ that Eq.~(\ref{cor:eq2333}) indeed holds true provided that the series (B.3) is used as the generating function. Equation (B.3) leads to suggestion about validity of (\ref{cor:eq2333}) at arbitrary $l$.

\end{document}